\documentclass[aps,prd,10pt,twocolumn,superscriptaddress,nofootinbib,longbibliography,floatfix]{revtex4-2}

\usepackage{amsmath,amssymb}
\usepackage{graphicx}
\usepackage{xcolor}
\usepackage[colorlinks=true,linkcolor=blue,citecolor=blue,urlcolor=blue]{hyperref}
\usepackage{bm}

\begin{document}

\title{Pole parameters of the $f_2(1270)$ and $f_2'(1525)$ from a coupled-channel
analysis of the radiative $J/\psi$ multipoles}

\author{Igor Danilkin}
\affiliation{Institut f\"ur Kernphysik and PRISMA$^+$ Cluster of Excellence,
Johannes Gutenberg Universit\"at, 55099 Mainz, Germany}

\author{David Baron}
\affiliation{Departament de F\'{i}sica Te\`{o}rica, Instituto de F\'{i}sica
Corpuscular, Universitat de Val\`encia - Consejo Superior de Investigaciones
Cient\'{i}ficas, Parc Cient\'{i}fic, Catedr\'{a}tico Jos\'{e} Beltr\'{a}n 2,
E-46980 Paterna, Valencia, Spain}

\author{Emilie Passemar}
\affiliation{Physics Department, Indiana University, Bloomington,
Indiana 47405, USA}
\affiliation{Departament de F\'{i}sica Te\`{o}rica, Instituto de F\'{i}sica
Corpuscular, Universitat de Val\`encia - Consejo Superior de Investigaciones
Cient\'{i}ficas, Parc Cient\'{i}fic, Catedr\'{a}tico Jos\'{e} Beltr\'{a}n 2,
E-46980 Paterna, Valencia, Spain}

\author{Marc Vanderhaeghen}
\affiliation{Institut f\"ur Kernphysik and PRISMA$^+$ Cluster of Excellence,
Johannes Gutenberg Universit\"at, 55099 Mainz, Germany}

\date{\today}

\begin{abstract}
We determine the pole parameters of $f_2(1270)$ and $f_2'(1525)$ from the combined fit to the isoscalar $D$-wave $\pi\pi$ phase shift, the $\pi\pi\to K\bar K$ modulus, and the BESIII $J/\psi\to\gamma\pi^0\pi^0$ and $\gamma K_SK_S$ data: all three $2^{++}$ multipoles $E1$, $M2$ and $E3$, and the $M2$-$E1$ phases. The hadronic amplitude is a two-channel $K$-matrix with two poles and all parameters free, so that no resonance information is taken from the Particle Data Group.  The radiative decay amplitudes are written as a coupled-channel Muskhelishvili-Omn\`es representation, with the Omn\`es matrix generated by that hadronic amplitude.  The left-hand cuts, which carry the production dynamics, are approximated by a linear subtraction polynomial in the physical region.  The model-independent kinematic constraints among the $J=2$ helicity amplitudes are implemented explicitly.  The fit describes all $358$ points with $\chi^2/N=1.29$ and gives $\sqrt{s_p}=(1267.5\pm1.3\pm2.6)-i\,\frac{1}{2}\,(175.5\pm2.4\pm7.4)\ \mathrm{MeV}$ for the $f_2(1270)$ and $(1514.8\pm2.8\pm1.5)-i\,\frac{1}{2}\,(80.4\pm3.4\pm4.8)\ \mathrm{MeV}$ for the $f_2'(1525)$, where the first uncertainty is statistical, from a parametric bootstrap with the data errors rescaled by $\sqrt{\chi^2/N}$ of each set, and the second systematic, from $17$ variations of the analysis.
\end{abstract}

\maketitle

\section{Introduction}
\label{sec:intro}

Below about $1.6\ \mathrm{GeV}$, the isoscalar $D$-wave system is governed by two resonances with nearly ideal flavour mixing.  The $f_2(1270)$ resonance couples mainly to $\pi\pi$, while $f_2'(1525)$ couples mainly to $K\bar K$. Lattice QCD~\cite{Briceno:2017qmb} finds the same pattern at $m_\pi\simeq391\ \mathrm{MeV}$: two $J^P=2^+$ resonances are found, the lighter decays dominantly to $\pi\pi$ and the heavier to $K\bar K$.  Being process independent, their pole positions are the quantities to determine.  The data are limited: with no meson beams available, the $\pi\pi$ phase shift~\cite{Hyams:1973zf,Protopopescu:1973sh} and the $\pi\pi\to K\bar K$ modulus~\cite{Longacre:1986fh} are both extracted from $\pi N\to\pi\pi N$, $\pi N\to\pi\pi\Delta$ and $\pi N\to K\bar K N$ under a one-pion-exchange assumption~\cite{Chew:1958wd,Pelaez:2015qba}.  The former constrains mainly the $f_2(1270)$, while the $f_2'(1525)$ is accessible through the $\pi\pi\to K\bar K$ transition amplitude and through other production reactions~\cite{CrystalBarrel:2019zqh,BESIII:2025dfb}, among them the radiative $J/\psi$ decays~\cite{BESIII:2015rug,BESIII:2018ubj} used here.

The BESIII data on $J/\psi\to\gamma\pi^0\pi^0$~\cite{BESIII:2015rug} and $\gamma K_SK_S$~\cite{BESIII:2018ubj} offer a good testing ground for the $f_2(1270)$ and $f_2'(1525)$ poles, since a mass-independent partial-wave analysis isolates the $2^{++}$ contribution bin by bin.  BESIII report the intensities of the three $2^{++}$ multipoles $E1$, $M2$ and $E3$ and their phases relative to $E1$.  These data have already been used to extract pole parameters. The Joint Physics Analysis Center (JPAC)~\cite{Rodas:2021tyb} fitted only the $0^{++}$ and $2^{++}$ $E1$ waves in the $N/D$ formalism up to $2.5\ \mathrm{GeV}$, using many parametrisations to control the model dependence. The $\pi\pi\to\pi\pi$ and $\pi\pi\to K\bar K$ scattering data were not included.  In Ref.~\cite{Danilkin:2025kyo} we set up a different framework, which we use here.  In the two-channel $\pi\pi/K\bar K$ approximation, the hadronic $D$ wave rescattering is described by a $K$-matrix, and the radiative decay amplitudes are then obtained from a coupled-channel Muskhelishvili-Omn\`es (MO) representation. The Omn\`es matrix is built from that $K$-matrix. In Ref.~\cite{Danilkin:2025kyo} the $K$-matrix masses/widths were taken from the PDG 2024~\cite{ParticleDataGroup:2024cfk}, and its couplings fitted to the $\pi\pi\to K\bar K$ modulus~\cite{Longacre:1986fh}, while the $2^{++}$ $E1$ data tested the construction as a whole.  The resonance parameters were therefore an input rather than a result.

Here we extend that analysis in three directions.  First, all $K$-matrix parameters are now free and are fitted together with the production subtraction constants.  The poles are therefore determined by the same fit that describes the data, with no PDG input on either resonance.  The Omn\`es matrix is rebuilt from the current $K$-matrix at every step. Second, we include the $M2$ and $E3$ spectra and the measured $M2$-$E1$ phases in the fit.  Finally, we implement the model-independent kinematic constraints among the $J=2$ helicity amplitudes~\cite{Danilkin:2018qfn} in both channels, as was done for $\pi^0\pi^0$ in Ref.~\cite{Hoid:2026ehk}.  The output is a pair of pole positions determined by scattering and radiative data together, which we compare with the other available extractions. The updated coupled-channel Omn\`es matrix is a second output and a useful building block for other dispersive analyses, see e.g.\ Refs.~\cite{TarrusCastella:2021pld,Cao:2024zlf,Yan:2026oil}.

\section{Kinematics and multipole decomposition}
\label{sec:kin}

We consider the radiative decays
\begin{equation*}
  J/\psi(q)\to\gamma(k)\,P(p_1)\,\bar P(p_2),\qquad P\bar P=\pi\pi,\ K\bar K,
\end{equation*}
with $s=(p_1+p_2)^2$ being the squared invariant mass of the meson pair and $q^2$ the squared $J/\psi$ mass.  The meson pair is treated in the isospin basis, with the two isoscalar channels labelled
\begin{equation*}
  a=1\equiv\pi\pi,\qquad a=2\equiv K\bar K.
\end{equation*}
We work in the isospin limit, taking $m_\pi=139.57\ \mathrm{MeV}$ and $m_K=495.64\ \mathrm{MeV}$. The mass of a single meson in channel $a$ is denoted by $m_a$.

For a real photon, parity leaves three independent helicity amplitudes $H^{(a)}_{\lambda_1\lambda_2}(s,\theta)$, corresponding to the helicity combinations $(\lambda_1\lambda_2)=(+-)$, $(+0)$ and $(++)$. Here $\lambda_1$ and $\lambda_2$ denote the helicities of the photon and the $J/\psi$ meson, respectively, and $\theta$ is the angle between the photon and one of the mesons in the rest frame of the pair. Their partial-wave expansion~\cite{Jacob:1959at} reads
\begin{equation}
  H^{(a)}_{\lambda_1\lambda_2}(s,\theta)=\mathcal{N}_a\sum_J(2J+1)\,
    h^{(a)}_{J,\lambda_1\lambda_2}(s)\,
    d^{J}_{\lambda_2-\lambda_1,0}(\theta),
  \label{eq:pw}
\end{equation}
being $d^{J}$ the Wigner rotation function and
$\mathcal{N}_1=1$, $\mathcal{N}_2=1/\sqrt2$. The normalization in the kaon channel ensures that both channels obey the same unitarity condition~\cite{Garcia-Martin:2010kyn,Danilkin:2020pak}.

Isospin relates these isoscalar amplitudes to the observed final states.  The short-distance source $c\bar c\to\gamma gg$ produces an isoscalar meson pair, and isospin-violating contributions to $J/\psi$ decays are electromagnetic and small (see the discussion in Refs.~\cite{Hoid:2026ehk,Rodas:2021tyb}).  Assuming isospin conservation, it holds
\begin{equation*}
  H^{\pi^0\pi^0}_{\lambda_1\lambda_2}
    =-\frac{1}{\sqrt3}\,H^{(1)}_{\lambda_1\lambda_2},\qquad
  H^{K^0\bar K^0}_{\lambda_1\lambda_2}
    =-\frac{1}{\sqrt2}\,H^{(2)}_{\lambda_1\lambda_2},
\end{equation*}
The $K_SK_S$ amplitude is identified with the $K^0\bar K^0$ one for even $J$ ($CP$ violation neglected).  Both physical final states, $P\bar P=\pi^0\pi^0$ and $K_SK_S$, consist of identical particles and carry the symmetry factor $1/2$ in the rate.  With it included, the decay distribution is
\begin{align}
  \frac{d\Gamma_{J/\psi\to\gamma P\bar P}}{d\sqrt s\,d\cos\theta}
  &=\frac{(q^{2}-s)\sqrt{s-4m_{P}^{2}}}{768\pi^{3}q^{3}}
   \sum_{\lambda_1\lambda_2}\left|H^{P\bar P}_{\lambda_1\lambda_2}(s,\theta)\right|^{2},
  \label{eq:width}\\
  q \equiv M_{J/\psi}\nonumber
\end{align}
where $m_P$ is the mass of the physical meson and the sum runs over the three helicity combinations, and the $J/\psi$ polarisations are averaged.

Bose symmetry restricts the sum in Eq.~\eqref{eq:pw} to even $J$, and we work
throughout with the three $J=2$ waves $h^{(a)}_{2,\lambda_1\lambda_2}$,
corresponding to the $2^{++}$ sector of the BESIII
analyses~\cite{BESIII:2015rug,BESIII:2018ubj}.  These partial-wave helicity amplitudes satisfy model-independent kinematic constraints that follow from the Bardeen-Tung-Tarrach decomposition of the hadronic tensor~\cite{Bardeen:1968ebo,Tarrach:1975tu}. In this decomposition, the hadronic tensor is written in terms of gauge-invariant tensor structures whose scalar coefficients are free of kinematic singularities and constraints. The resulting relations factor out the required kinematic zeros at the two-body threshold $s=4m_a^2$ and the soft-photon point $s=q^2$. The soft-photon behavior is governed by Low's theorem~\cite{Low:1958sn}. These relations were derived for $\gamma\gamma^{*}\to\pi\pi$ in Refs.~\cite{Danilkin:2018qfn,Danilkin:2019opj}\footnote{An equivalent formulation is given in Ref.~\cite{Hoferichter:2019nlq}.} and later adapted to the decay kinematics in Ref.~\cite{Hoid:2026ehk}. They read
\begin{align}
  h^{(a)}_{2,+-}&=T_1^{(a)}h^{(a)}_1,\nonumber\\
  h^{(a)}_{2,+-}-\sqrt{\tfrac{2s}{q^2}}\,h^{(a)}_{2,+0}
    &=T_2^{(a)}h^{(a)}_2,\nonumber\\
  h^{(a)}_{2,+-}-2\sqrt{\tfrac{2s}{q^2}}\,h^{(a)}_{2,+0}
    +\tfrac{\sqrt6\,s}{q^2}h^{(a)}_{2,++}&=T_3^{(a)}h^{(a)}_3,
  \label{eq:kinconstraints}
\end{align}
with
\begin{equation}
  T_i^{(a)}(s)=\left(s-4m_a^2\right)\left(s-q^2\right)^i ,
  \qquad i=1,2,3,
  \label{eq:Ti}
\end{equation}
and $h^{(a)}_i(s)$ regular at both kinematic points.

The radiative multipoles used in charmonium radiative transitions~\cite{BESIII:2015rug,BESIII:2018ubj} are related to the helicity partial waves  by the orthogonal combinations~\cite{Karl:1975jp,Olsson:1986aw,Sebastian:1992xq}
\begin{equation}
  \begin{pmatrix}h^{(a)}_{2,E1}\\h^{(a)}_{2,M2}\\h^{(a)}_{2,E3}\end{pmatrix}
  =\begin{pmatrix}
     \sqrt{3/5} & \sqrt{3/10} & 1/\sqrt{10}\\
     -1/\sqrt3 & 1/\sqrt6 & 1/\sqrt2\\
     1/\sqrt{15} & -2\sqrt{2/15} & \sqrt{2/5}
   \end{pmatrix}
  \begin{pmatrix}h^{(a)}_{2,+-}\\h^{(a)}_{2,+0}\\h^{(a)}_{2,++}\end{pmatrix}.
  \label{eq:rot}
\end{equation}
Inverting Eq.~\eqref{eq:kinconstraints} and substituting the result into Eq.~\eqref{eq:rot} gives $E1$, $M2$ and $E3$ in each channel in terms of the three regular functions $h^{(a)}_i$.  The dynamics enters only through these functions, which are the subject of the next section.

\section{Amplitude}
\label{sec:model}

\subsection{Coupled-channel $K$-matrix}
\label{sec:kmat}

We parametrize the hadronic amplitude with a two-channel $K$-matrix.  The isoscalar $D$-wave
$t$-matrix reads
\begin{equation}
  \bm{t}^{-1}(s)=\bm{K}^{-1}(s)-i\bm{\rho}(s),\qquad
  \rho_a(s)=p_a^{2J}\,\frac{2p_a}{\sqrt s},
  \label{eq:tmat}
\end{equation}
where, $\bm{\rho}=\text{diag}\left(\rho_1(s)\theta(s-4m_\pi^2),\;\rho_2(s)\theta(s-4m_K^2)\right)$, $J=2$, $p_a(s)=\tfrac12\sqrt{s-4m_a^2}$. The matrix $\bm{K}$ is real and symmetric on the real axis, ensuring two-channel unitarity. The $K$-matrix contains the two resonances only,
\begin{equation}
  K_{ab}(s)=\sum_{R=1}^{2}
    \frac{\tilde g_{R,a}(s)\,\tilde g_{R,b}(s)}{m_R^2-s},
  \label{eq:kmat}
\end{equation}
with Blatt-Weisskopf factors
\begin{align}
  \tilde g_{R,a}(s)&=g_{R,a}\,
    \frac{F_2\!\left(p_a(s)R_{\rm int}\right)}
         {F_2\!\left(p_a(m_R^2)R_{\rm int}\right)},\nonumber\\
  F_2(x)&=\frac{1}{\sqrt{x^4+3x^2+9}} ,
  \label{eq:BW}
\end{align}
normalised to unity at the resonance mass.  The interaction radius $R_{\rm int}$ is fitted.  The Omn\`es problem of the next subsection is solved without subtractions, which requires the eigenphases to obey~\cite{Muskhelishvili:1953,Omnes:1958hv,Moussallam:1999aq}
\begin{equation}
  \sum_a\delta_a(\infty)=2\pi.
  \label{eq:asym}
\end{equation}
The $K$-matrix is constructed to comply with it: with two pole terms and no background, both eigenphases tend to $\pi$ at infinity holding Eq.~\eqref{eq:asym} automatically.

Within the two-channel model the $\pi\pi$ amplitude is elastic below the $K\bar K$ threshold, and the residues of the poles are shared between $\pi\pi$ and $K\bar K$ alone.  The fitted couplings therefore compensate the omitted $4\pi$ and $\eta\eta$ channels of the two resonances.  This is the same simplification used in Ref.~\cite{Danilkin:2025kyo}, and its consequences are discussed in Sec.~\ref{sec:err}.

We have tested whether the data require a smooth non-resonant term. To this end, we add to the $K$-matrix the term $\Delta K_{ab}(s)=C_{ab}\,b_a(s)\,b_b(s)/(s_B-s)$, where $C$ is a real symmetric matrix with three independent entries and $s_B=-0.5\ \mathrm{GeV}^2$ is fixed. Here $b_a(s)=F_2(p_a(s)R_{\rm int})/F_2(p_a(s_0)R_{\rm int})$, with $s_0=1\ \mathrm{GeV}^2$, is the barrier factor of Eq.~\eqref{eq:BW}. It falls off like the resonance terms, so that Eq.~\eqref{eq:asym} still holds, and provides a standard effective description of crossed-channel exchange. With all other inputs unchanged, the three entries of $C$ lower the $\chi^2$ by $6.0$ out of a total of $461.3$. The pole masses shift by at most $1.6\ \mathrm{MeV}$ and the widths by at most $1.2\ \mathrm{MeV}$. We therefore retain the two-pole $K$-matrix of Eq.~\eqref{eq:kmat} and include these shifts in the systematic uncertainty (see Sec.~\ref{sec:err}).

\subsection{Omn\`es matrix}
\label{sec:omnes}

The radiative decays are described with the help of the coupled-channel Omn\`es
matrix, which solves the homogeneous MO
problem~\cite{Garcia-Martin:2010kyn,Moussallam:1999aq}.  We use it in unsubtracted
form,
\begin{align}
  \bm{\Omega}(s)&=\frac{1}{\pi}\int_{s_{\rm th_1}}^{\infty}\!{\rm d}s'\,
         \frac{{\rm Im}\,\bm{\Omega}(s')}{s'-s},\nonumber\\
  {\rm Im}\,\bm{\Omega}(s)&=\bm{t}^{*}(s)\,\bm{\rho}(s)\,\bm{\Omega}(s),
  \label{eq:omnesDR}
\end{align}
with $s_{\rm th_1}=4m_\pi^2$.  Substituting the imaginary
part gives a real principal-value integral equation for ${\rm Re}\,\bm{\Omega}$,
\begin{align}
  {\rm Re}\,\bm{\Omega}(s)&=\frac{1}{\pi}\,{\rm P.V}\!\!\int_{s_{\rm th_1}}^{\infty}\!
     {\rm d}s'\,\frac{\bm{X}(s')\,{\rm Re}\,\bm{\Omega}(s')}{s'-s},\nonumber\\
  \bm{X}(s)&=\left[\bm{1}-i\,\bm{t}^{*}(s)\bm{\rho}(s)\right]^{-1}
             \bm{t}^{*}(s)\,\bm{\rho}(s),
  \label{eq:omnesRe}
\end{align}
which is discretised on a Gauss-Legendre mesh using the principal-value kernel of Ref.~\cite{Moussallam:1999aq} and solved by singular-value decomposition~\cite{Golub:2013}.  For a given $\bm{t}$, its solutions form a two-dimensional space when Eq.~\eqref{eq:asym} holds. The normalisation $\bm{\Omega}(0)=\bm{1}$ then fixes $\bm{\Omega}$ uniquely. The Omn\`es matrix $\bm{\Omega}$ is rebuilt from the current $\bm{t}$-matrix at every step of the minimisation.

\subsection{The three $2^{++}$ multipoles}
\label{sec:mult}

\begin{figure}[!tbp]
\includegraphics[width=0.98\columnwidth]{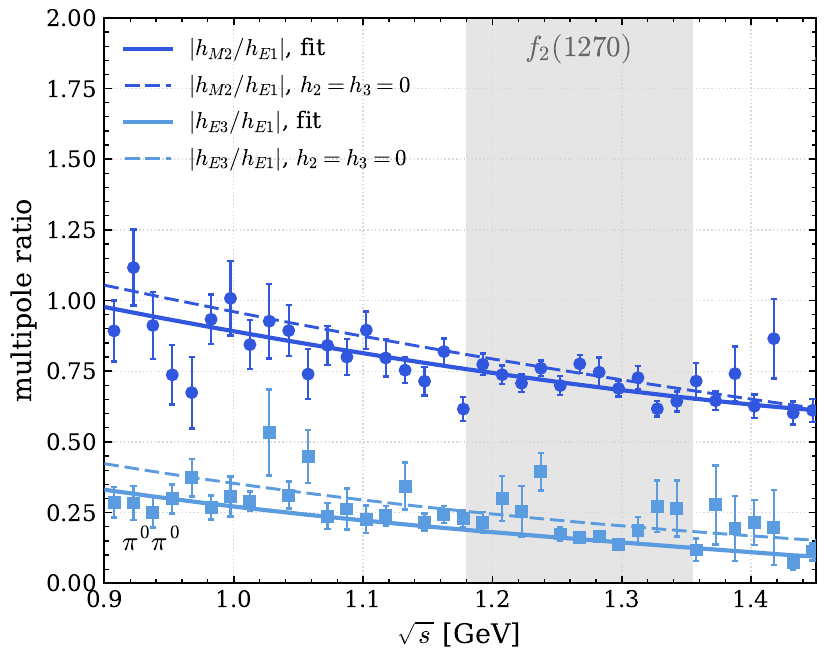}
\caption{The $\pi^0\pi^0$ multipole ratios.  Dashed: the limit $h_2=h_3=0$ (see Eq.~\eqref{eq:kinratio}), which contains no fitted quantity. Solid: the full amplitude of the adopted fit, with $h_2$ and $h_3$ included. Points: $\sqrt{I_{M2}/I_{E1}}$ (circles) and $\sqrt{I_{E3}/I_{E1}}$ (squares) formed bin by bin from the BESIII $J/\psi\to\gamma\pi^0\pi^0$ multipole intensities~\cite{BESIII:2015rug}.  Above $1.45\ \mathrm{GeV}$, the $E1$ intensity is small and the measured ratios are noisy.}
\label{fig:hierarchy}
\end{figure} 
The regular functions $h^{(a)}_i$ introduced in Sec.~\ref{sec:kin} are the production amplitudes of the two channels, with the kinematic factors of Eq.~\eqref{eq:Ti} removed.  Their right-hand cut is generated by the $\pi\pi/K\bar K$ rescattering and is carried by the Omn\`es matrix. Therefore, they obey the MO representation~\cite{Garcia-Martin:2010kyn,Moussallam:2013una,Danilkin:2018qfn}. With $\bm{h}_i=(h^{(1)}_i,h^{(2)}_i)^{T}$ it holds
\begin{align*}
  \bm{h}_i(s)&=\bm{\Omega}(s)\left[\frac{1}{\pi}\int_{L}{\rm d}s'\,
     \frac{\bm{\Omega}^{-1}(s')\,{\rm Im}\,\bm{h}_i(s')}{s'-s}\right]\\
  &\approx\bm{\Omega}(s)\,\bm{P}_i(s),
\end{align*}
where $L$ denotes the left-hand cuts of the crossed-channel exchanges.  In the physical region, the left-hand-cut integral is a slowly varying function. It can therefore be approximated by a linear polynomial $\bm{P}_i(s)$.  In components,
\begin{equation}
  h^{(a)}_i(s)=\sum_b\Omega_{ab}(s)\left(\alpha_i^b+\beta_i^b s\right),
  \qquad i=1,2,3.
  \label{eq:Hbasis}
\end{equation}
The same approximation was used in Ref.~\cite{Danilkin:2025kyo}. In Ref.~\cite{Hoid:2026ehk} the authors explicitly estimated the leading crossed-channel exchanges, $\rho$ and $b_1$, and found their contributions too small to affect the fits of the $S$-wave $\pi^0\pi^0$ data. The production sector is thus twelve real constants. The fit is nested: the minimiser varies the seven $K$-matrix parameters, and at each step the twelve constants are refitted to the radiative data by an inner least-squares fit. As a test of the linear approximation, we add a real quadratic term $\gamma_i^b s^2$ to all six polynomials, with all other inputs unchanged. The six additional constants lower the $\chi^2$ by about $50$, much more than the background term of the $K$-matrix in Sec.~\ref{sec:kmat}. We keep the linear polynomials and include the resulting pole shifts, listed in Appendix~\ref{app:syst}, in the systematic uncertainty (see Sec.~\ref{sec:err}).

Much of the observed multipole hierarchy follows from Eq.~\eqref{eq:kinconstraints} when $h_1$ dominates over $h_2$ and $h_3$. With $h_2=h_3=0$ the three helicity amplitudes are proportional to a single function,
\begin{equation}
  h_{2,+0}=\sqrt{\frac{q^2}{2s}}\,h_{2,+-},\qquad
  h_{2,++}=\frac{q^2}{\sqrt6\,s}\,h_{2,+-}.
  \label{eq:hellimit}
\end{equation}
The rotation~\eqref{eq:rot} then gives not a pure $E1$ amplitude but $M2$ and $E3$ as multiples of $E1$,
\begin{align}
  \left.\frac{h_{M2}}{h_{E1}}\right|_{h_2=h_3=0}&=\sqrt5\,\frac{x+\sqrt x-2}{x+3\sqrt x+6},\nonumber\\
  \left.\frac{h_{E3}}{h_{E1}}\right|_{h_2=h_3=0}&=\frac{2(\sqrt x-1)^2}{x+3\sqrt x+6},\qquad
  x=\frac{q^2}{s}.
  \label{eq:kinratio}
\end{align}
These ratios depend only on $s$ and $q^2$ and involve no fitted quantity, so that they provide a parameter-free test of this limit. Both ratios are real and positive for all $s<q^2$, therefore in this limit the $M2$-$E1$ and $E3$-$E1$ phase differences vanish identically\footnote{In a single-channel approximation, Watson's theorem~\cite{Watson:1954uc} gives every multipole the phase $\delta_1$, but only modulo $\pi$.  It therefore does not guarantee that the phase differences vanish, since they could equally be
$\pm \pi$.},
\begin{align}
  \big(\arg h_{M2}-\arg h_{E1}\big)\big|_{h_2=h_3=0}&=0,\nonumber\\
  \big(\arg h_{E3}-\arg h_{E1}\big)\big|_{h_2=h_3=0}&=0.
  \label{eq:zerophase}
\end{align}
Near the soft-photon point, the amplitude ratios reduce to the familiar multipole hierarchy, $h_{M2}/h_{E1}\propto E_\gamma$ and $h_{E3}/h_{E1}\propto E_\gamma^2$.  At the $f_2(1270)$ peak Eq.~\eqref{eq:kinratio} gives $|h_{M2}/h_{E1}|=0.74$ and $|h_{E3}/h_{E1}|=0.21$, against $0.78\pm0.03$ and $0.16\pm0.01$ for the square roots of the measured intensity ratios in $\pi^0\pi^0$. At the $f_2'(1525)$ peak it gives $0.57$ and $0.13$, against $0.51\pm0.05$ and $0.10\pm0.07$ in $K_SK_S$.  Figure~\ref{fig:hierarchy} shows the comparison in $\pi^0\pi^0$ between $0.9$ and $1.45\ \mathrm{GeV}$, where the measured ratios are precise.  

The observation that the parameter-free relations resulting from the limit $h_2=h_3=0$ are a remarkably good approximation for $J/\psi$ is worth discussing in more detail. This limit corresponds to a single production vertex,
\begin{equation}
  \mathcal{M}=g\,\epsilon^{*}_{\mu\nu}\,F^{\mu\alpha}\,\psi^{\nu}{}_{\alpha},
  \label{eq:vertex}
\end{equation}
where $\epsilon_{\mu\nu}$ is the polarisation tensor of the $J=2$ meson pair, $F^{\mu\nu}$ and $\psi^{\mu\nu}$ are the field strengths of the photon and the $J/\psi$, and $g$ is a coupling constant. The general vertex has three independent couplings, one for each helicity amplitude, and Eq.~\eqref{eq:vertex} is the one of lowest dimension. The other two contain two more derivatives and generate $h_2$ and $h_3$. For $q^2=0$ (the case of two real photons) the coupling~\eqref{eq:vertex} results in helicity-2 only~\cite{Drechsel:1999rf,Lu:2020qeo}, which is the helicity-2 dominance seen in $\gamma\gamma\to f_2(1270)$~\cite{Belle:2008bmg,Dai:2014zta}. At $q^2=M_{J/\psi}^2$ the factors $\sqrt{q^2/2s}$ and $q^2/(\sqrt6\,s)$ of Eq.~\eqref{eq:hellimit} turn the same coupling into the helicity-0 dominance expected for two-gluon production of a light tensor meson~\cite{Krammer:1978qp,Korner:1982vg}. In this limit helicity-0 carries $60\%$ of the rate at the $f_2(1270)$ peak. At the $\Upsilon(1S)$ mass this fraction rises to $95\%$, in line with the helicity-0 dominance observed in $\Upsilon(1S)\to\gamma f_2(1270)$~\cite{CLEO:2005koa,BaBar:2018uqa}.

In the fit, $h_1$ indeed dominates the intensities, and $h_2$ and $h_3$ enter the moduli only as corrections (see Table~\ref{tab:prod}).  For $h_3$, the linear production polynomial could be replaced by a constant.  This replacement is one of the variations used to estimate the systematic uncertainties in Appendix~\ref{app:syst}. All three functions are nevertheless kept free. Since the phase differences vanish in the $h_2=h_3=0$ limit, the relative phases between the multipoles are carried entirely by the $h_2$ and $h_3$ admixtures, and the sign with which they enter can be resolved by the measured phase differences (see Sec.~\ref{sec:phases}).

\section{Data and fit}
\label{sec:fit}

\begin{table}[!tbp]
\caption{Data sets included in the fit and the fit quality for each set. $N$ is the number of data set points, and $\chi^2/N$ denotes the $\chi^2$ per point, rather than per degree of freedom. The last column gives the scale factor $S=\sqrt{\chi^2/N}$ used to enlarge the errors of each set before generating the bootstrap samples described in Sec.~\ref{sec:err}, following the PDG prescription. Sets described within their errors retain $S=1$. For both final states, only BESIII solution~1 is used.}
\label{tab:chi2}
\begin{ruledtabular}
\begin{tabular}{lrrrr}
input & $N$ & $\chi^2$ & $\chi^2/N$ & $S$ \\ \hline
$\delta_1$, CFD~\cite{Garcia-Martin:2011iqs}, $\sqrt s<0.9\ \mathrm{GeV}$ & 12 & $6.3$ & $0.53$ & $1.00$ \\
$\delta_1$, Hyams~\cite{Hyams:1973zf}, $0.9$-$1.5\ \mathrm{GeV}$ & 30 & $59.3$ & $1.98$ & $1.41$ \\
$|t_{12}|$, Brookhaven~II~\cite{Longacre:1986fh} & 16 & $17.4$ & $1.09$ & $1.04$ \\
$J/\psi\to\gamma\pi^0\pi^0$, $E1$~\cite{BESIII:2015rug} & 47 & $59.4$ & $1.26$ & $1.12$ \\
$J/\psi\to\gamma\pi^0\pi^0$, $M2$~\cite{BESIII:2015rug} & 47 & $80.3$ & $1.71$ & $1.31$ \\
$J/\psi\to\gamma\pi^0\pi^0$, $E3$~\cite{BESIII:2015rug} & 47 & $49.1$ & $1.04$ & $1.02$ \\
$J/\psi\to\gamma K_SK_S$, $E1$~\cite{BESIII:2018ubj} & 41 & $39.4$ & $0.96$ & $1.00$ \\
$J/\psi\to\gamma K_SK_S$, $M2$~\cite{BESIII:2018ubj} & 41 & $48.8$ & $1.19$ & $1.09$ \\
$J/\psi\to\gamma K_SK_S$, $E3$~\cite{BESIII:2018ubj} & 41 & $21.5$ & $0.52$ & $1.00$ \\
$\arg h_{M2}-\arg h_{E1}$, $\pi^0\pi^0$~\cite{BESIII:2015rug} & 29 & $60.3$ & $2.08$ & $1.44$ \\
$\arg h_{M2}-\arg h_{E1}$, $K_SK_S$~\cite{BESIII:2018ubj} & 7 & $19.5$ & $2.79$ & $1.67$ \\
\hline
\textbf{total} & 358 & $\mathbf{461.3}$ & $\mathbf{1.29}$ & \\
\end{tabular}
\end{ruledtabular}
\end{table}

\begin{figure*}[!tbp]
\includegraphics[width=0.98\textwidth]{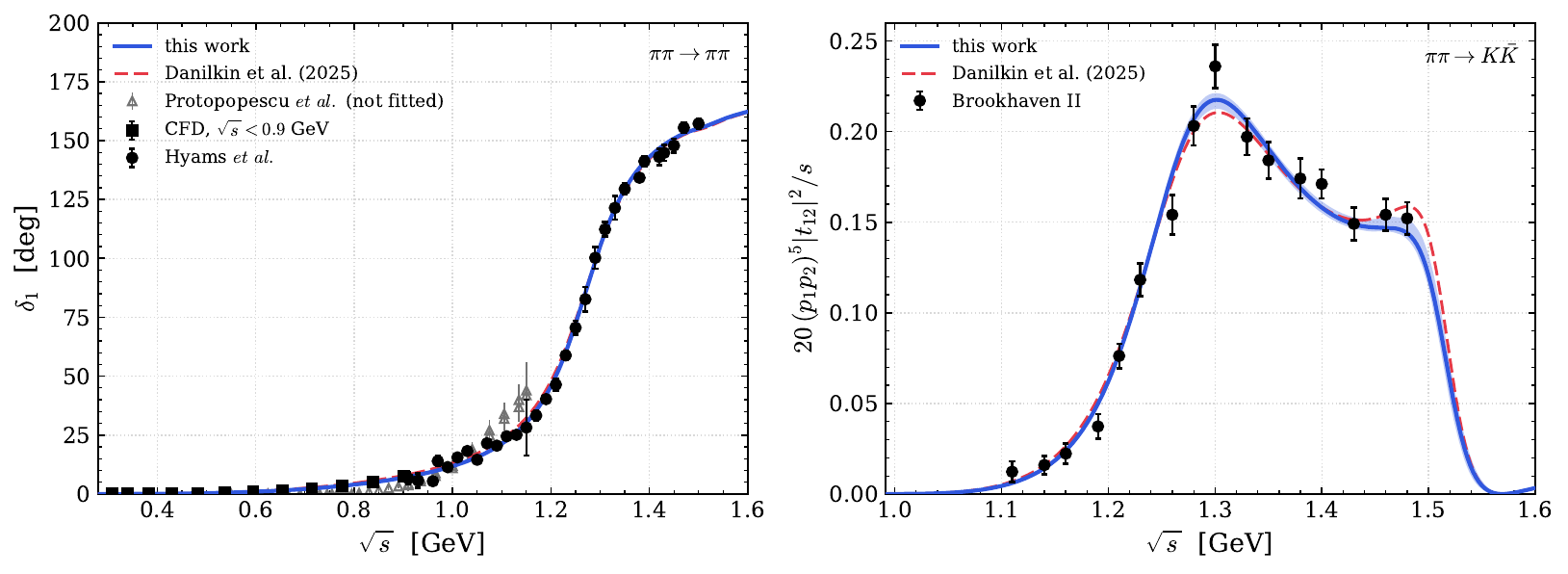}
\caption{The hadronic sector.  Left: the isoscalar $D$-wave phase shift
against the CFD points~\cite{Garcia-Martin:2011iqs,Pelaez:2024uav} used below $0.9\ \mathrm{GeV}$ and
the Hyams points~\cite{Hyams:1973zf} above.  The Protopopescu set~\cite{Protopopescu:1973sh} is
shown in grey and not fitted.  Right: the $\pi\pi\to K\bar K$ modulus against
Brookhaven~II~\cite{Longacre:1986fh}.  The band is the $68\%$ interval of the
bootstrap ensemble of Sec.~\ref{sec:err}.  Dashed: the solution of
Ref.~\cite{Danilkin:2025kyo}.}
\label{fig:hadronic}
\end{figure*}

The fit includes eleven data sets, listed with their sizes in Table~\ref{tab:chi2}. Below $0.9\ \mathrm{GeV}$, we use the isoscalar $D$-wave phase shift taken from the CFD parametrisation of Refs.~\cite{Garcia-Martin:2011iqs,Pelaez:2024uav}, sampled at twelve points, each assigned a $5\%$ uncertainty, the size of the quoted uncertainties of the $D$-wave threshold parameters~\cite{Garcia-Martin:2011iqs}.  Its $D$ wave is a fit to the $\pi N$ data, constrained by forward dispersion relations. We use it in place of the raw data, as was done in Ref.~\cite{Pelaez:2025jrn}.  Above $0.9\ \mathrm{GeV}$, we use the Hyams data~\cite{Hyams:1973zf} up to $1.5\ \mathrm{GeV}$.  The Protopopescu set~\cite{Protopopescu:1973sh} is shown in Fig.~\ref{fig:hadronic} but not fitted, since it is inconsistent with Hyams in the overlap region at the level of its own errors.  The $\pi\pi\to K\bar K$ modulus is the Brookhaven~II measurement~\cite{Longacre:1986fh}.  We do not use the $D$-wave inelasticity measured by Hyams \emph{et al.}: below the $K\bar K$ threshold, the model is elastic, while above it the measured inelasticity includes contributions from the $4\pi$ channel, omitted by the model.

The radiative data are the six $2^{++}$ multipole spectra of the BESIII mass-independent analyses of $J/\psi\to\gamma\pi^0\pi^0$~\cite{BESIII:2015rug} and $J/\psi\to\gamma K_SK_S$~\cite{BESIII:2018ubj} for $0.9\le\sqrt s\le1.6\ \mathrm{GeV}$. Below $0.9\ \mathrm{GeV}$ the $2^{++}$ signal is weak, and the $D$-wave amplitude is fixed there by the CFD phase shift.  We fit the $J=2$ spectra $(1/\Gamma_f)\,d\Gamma^{X}_f/d\sqrt s$, with $X=E1,M2,E3$, normalised to the partial width $\Gamma_f=\Gamma_{J/\psi}\mathcal{B}_f$ of the final state (see Ref.~\cite{Danilkin:2025kyo} for details). The branching fractions $\mathcal{B}_f$ are taken from the PDG~\cite{ParticleDataGroup:2026mpi}.  Inserting Eq.~\eqref{eq:pw} into Eq.~\eqref{eq:width} and integrating over the angle, the $J=2$ term gives
\begin{align}
\frac{d\Gamma^{X}_{J/\psi\to\gamma\pi^{0}\pi^{0}}}{d\sqrt s}
   &=\frac{(q^{2}-s)\sqrt{s-4m_{\pi^0}^{2}}}{384\pi^{3}q^{3}}\,
     5\left|h^{\pi^{0}\pi^{0}}_{2,X}(s)\right|^{2},\nonumber\\
  \frac{d\Gamma^{X}_{J/\psi\to\gamma K_SK_S}}{d\sqrt s}
   &=\frac{(q^{2}-s)\sqrt{s-4m_{K_S}^{2}}}{384\pi^{3}q^{3}}\,
     5\left|h^{K^{0}\bar K^{0}}_{2,X}(s)\right|^{2},
  \label{eq:widths}
\end{align}
where the physical partial waves are $h^{\pi^0\pi^0}_{2,X}=-h^{(1)}_{2,X}/\sqrt3$ and $h^{K^0\bar K^0}_{2,X}=-h^{(2)}_{2,X}/2$, with the normalisation inherited from Eq.~\eqref{eq:pw}.  The overall normalisation is absorbed by the production constants.  They are common to both final states, so no separate normalisation factor is introduced for the $\pi^0\pi^0$ and $K_SK_S$
spectra.

Both BESIII analyses also tabulate the phases of $M2$ and $E3$ relative to $E1$, which is taken real. Two points have to be settled before these can be used, their convention and their credibility. As to the convention, BESIII build their multipoles from Clebsch-Gordan coefficients. We have verified that their three helicity combinations coincide, sign for sign, with the rows of Eq.~\eqref{eq:rot}. The tabulated phases are therefore directly comparable with ours, modulo $2\pi$.  The only freedom left is the reflection of a whole bin, which BESIII fix by a positive $0^{++}$-$E1$ phase. As to the credibility, the phase of an amplitude is undefined when the amplitude is compatible with zero, and BESIII caution that their errors hold only in the Gaussian limit. For phases with large uncertainties, this description may no longer be reliable. We therefore use a phase only in bins where both intensities entering the phase difference, $I_{E1}$ and $I_{M2}$ or $I_{E3}$, differ from zero by more than twice their respective errors, and where the quoted error of the phase is below $0.5$ rad.  This leaves $29$ bins of $\arg h_{M2}-\arg h_{E1}$ in $\pi^0\pi^0$, $7$ in $K_SK_S$, all on the $f_2'(1525)$ peak, and $14$ bins of $\arg h_{E3}-\arg h_{E1}$ in $\pi^0\pi^0$. Only one $K_SK_S$ $E3$ phase bin passes the cuts, and it is not used in the fit. The two $M2$ sets are fitted while the $E3$ set serves as a check (see Sec.~\ref{sec:phases}). The credibility cuts proposed above are varied among the systematic variations of Appendix~\ref{app:syst}.

The fit minimises the sum of the eleven $\chi^2$ of Table~\ref{tab:chi2} over the seven hadronic parameters, with the twelve production constants refitted at every step as described in Sec.~\ref{sec:mult}.  Since the intensities fix the constants only up to the relative signs of the three basis functions, the $\chi^2$ has several nearly degenerate minima corresponding to different sign patterns. To find the global minimum, the minimisation is done in steps: the production sector is started from every sign pattern, and the phase data select among the
resulting solutions (see Sec.~\ref{sec:phases}).

\section{Results}
\label{sec:res}

\begin{table}[!tbp]
\caption{$K$-matrix parameters from the adopted and $E1$-only fits, compared with Ref.~\cite{Danilkin:2025kyo}. The adopted fit includes the hadronic data, the six multipole spectra and the two $M2$-$E1$ phase difference sets. The $E1$-only fit includes the hadronic data and the two $E1$ spectra. Both fits use the same production sector, with all seven $K$-matrix parameters free. In the reference fit, the masses and widths were fixed to their PDG values, and $R_{\rm int}$ was also fixed. Bootstrap uncertainties (see Sec.~\ref{sec:err}) are given in parentheses.}
\label{tab:params}
\begin{ruledtabular}
\begin{tabular}{lrrr}
parameter & adopted fit & $E1$ only & Ref.~\cite{Danilkin:2025kyo} \\ \hline
$m_{1}$ [GeV] & $+1.276(1)$ & $+1.280(1)$ & $+1.275$ \\
$g_{1,\pi\pi}$ [GeV$^{-1}$] & $+1.211(8)$ & $+1.233(9)$ & $+1.246$ \\
$g_{1,K\bar K}$ [GeV$^{-1}$] & $+0.763(8)$ & $+0.763(8)$ & $+0.771$ \\
$m_{2}$ [GeV] & $+1.519(3)$ & $+1.514(3)$ & $+1.517$ \\
$g_{2,\pi\pi}$ [GeV$^{-1}$] & $+0.080(2)$ & $+0.078(3)$ & $+0.086$ \\
$g_{2,K\bar K}$ [GeV$^{-1}$] & $-1.210(26)$ & $-1.157(34)$ & $-1.139$ \\
$R_{\rm int}$ [GeV$^{-1}$] & $+3.360(42)$ & $+3.326(39)$ & $3.62$ (fixed) \\
\end{tabular}
\end{ruledtabular}
\end{table}
\begin{table}[!tbp]
\caption{The twelve production constants, defined in Eq.~\eqref{eq:Hbasis}, of the adopted fit. The coefficients $\alpha_i^b$ and $\beta_i^b$ are given in units of $10^{-3}\ \mathrm{GeV}^{-2-2i}$ and $10^{-3}\ \mathrm{GeV}^{-4-2i}$, respectively. The index $b$ labels the Omn\`es column. Bootstrap uncertainties (see Sec.~\ref{sec:err}) in parentheses.  $h_1$ dominates in both columns, and $h_2$ is significant in $K\bar K$ but not in $\pi\pi$.}
\label{tab:prod}
\begin{ruledtabular}
\begin{tabular}{lrr}
 & $b=\pi\pi$ & $b=K\bar K$ \\ \hline
$\alpha_{1}^{b}$ & $-0.353(21)$ & $-0.54(12)$ \\
$\beta_{1}^{b}$ & $+0.095(13)$ & $+0.149(51)$ \\
\hline
$\alpha_{2}^{b}$ & $+0.0018(22)$ & $+0.042(13)$ \\
$\beta_{2}^{b}$ & $-0.0018(13)$ & $-0.0212(56)$ \\
\hline
$\alpha_{3}^{b}$ & $+0.0004(3)$ & $-0.0047(23)$ \\
$\beta_{3}^{b}$ & $+0.0000(2)$ & $+0.0030(11)$ \\
\end{tabular}
\end{ruledtabular}
\end{table}

\begin{figure*}[!tbp]
\includegraphics[width=0.98\textwidth]{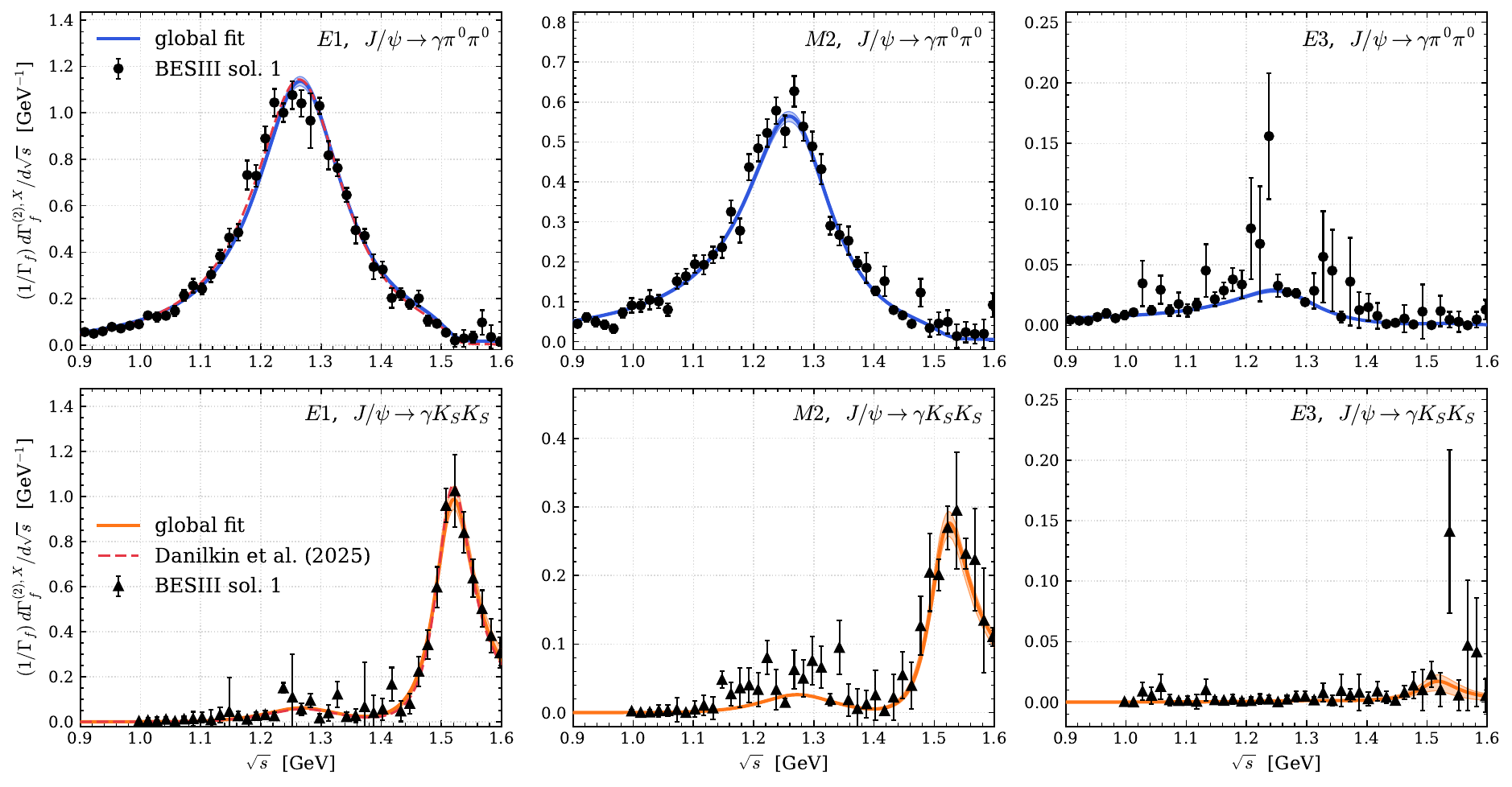}
\caption{The six BESIII $2^{++}$ multipole spectra (solution~1) against the
fit.  Top row $\pi^0\pi^0$~\cite{BESIII:2015rug}, bottom row $K_SK_S$~\cite{BESIII:2018ubj}.
The band is the $68\%$ interval of the bootstrap ensemble of Sec.~\ref{sec:err}.
Dashed: the $E1$ spectra of the solution of Ref.~\cite{Danilkin:2025kyo}.}
\label{fig:multipole}
\end{figure*}

\subsection{Description of the data}

Figures~\ref{fig:hadronic} and~\ref{fig:multipole} show the fit, and
Table~\ref{tab:chi2} its $\chi^2$ per set.  All eleven sets are described
together, with $\chi^2/N=1.29$ over $358$ points.  Of the three
hadronic inputs, only the Hyams phase shift is not reproduced within its
errors, at $1.98$ per point.  This tension is already present in the fit
to the $E1$ spectra alone, and therefore not caused by the $M2$ and $E3$ spectra or by the phases, and
we return to it in Sec.~\ref{sec:err}.  The two phase sets have the largest
$\chi^2/N$, $2.08$ and $2.79$, but their quoted errors hold only in
the Gaussian limit (see Sec.~\ref{sec:fit}).  The fitted $K$-matrix parameters are given in
Table~\ref{tab:params} and the production constants in Table~\ref{tab:prod}.
The interaction radius comes out at $R_{\rm int}=3.36\pm0.04\ \mathrm{GeV}^{-1}$,
somewhat below the $3.62\ \mathrm{GeV}^{-1}$ imposed in Ref.~\cite{Danilkin:2025kyo}.

\begin{figure*}[!tbp]
\includegraphics[width=0.84\textwidth]{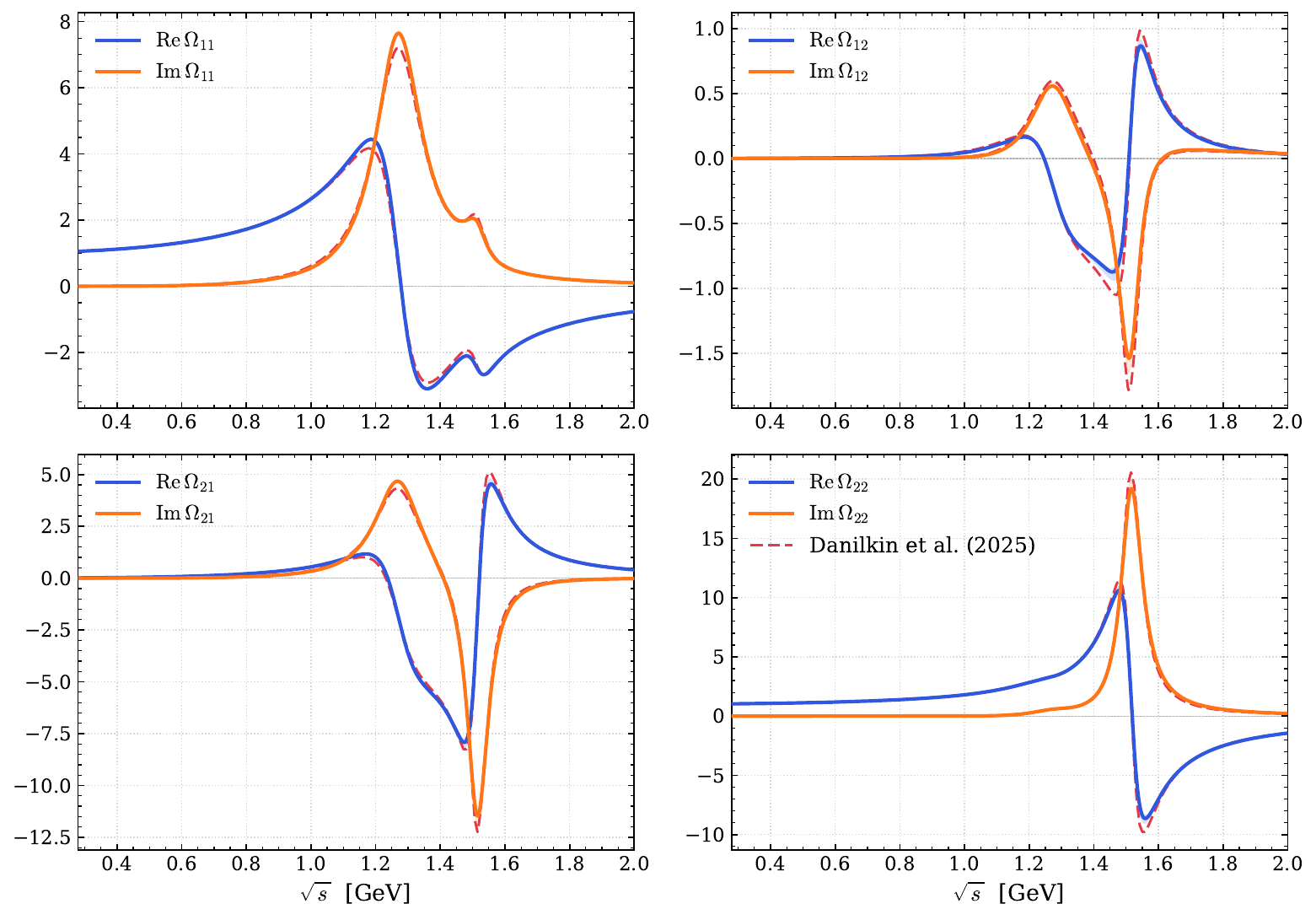}
\caption{The coupled-channel Omn\`es matrix elements of the fit.  Dashed: the matrix elements of
Ref.~\cite{Danilkin:2025kyo}.}
\label{fig:omnes}
\end{figure*}

The resulting Omn\`es matrix is shown in Fig.~\ref{fig:omnes}. Below the $f_2'(1525)$, it agrees with the matrix of Ref.~\cite{Danilkin:2025kyo}, whose $K$-matrix had the resonance parameters fixed to their PDG values. The two differ around the $f_2'(1525)$ resonance, whose width is $9\ \mathrm{MeV}$ smaller in Ref.~\cite{Danilkin:2025kyo}. The Omn\`es matrix can serve as dispersive input wherever an isoscalar $D$-wave
$\pi\pi/K\bar K$ final-state interaction is needed, see e.g.\
Refs.~\cite{TarrusCastella:2021pld,Cao:2024zlf,Yan:2026oil}.

\subsection{The relative phases}
\label{sec:phases}

\begin{figure*}[!tbp]
\includegraphics[width=0.84\textwidth]{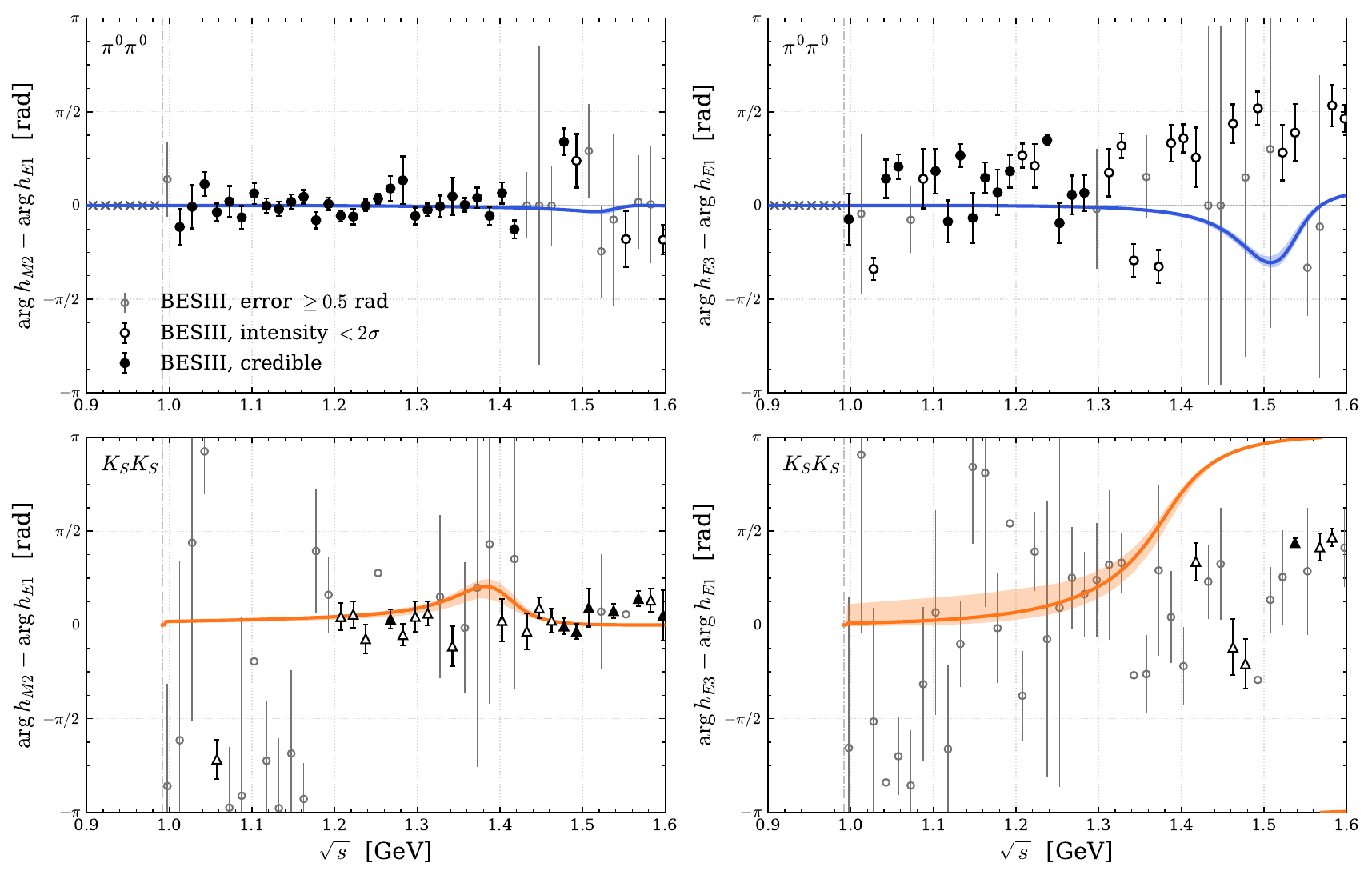}
\caption{The $J=2$ phase differences against BESIII~\cite{BESIII:2015rug,BESIII:2018ubj}. Filled: the credible bins, with a quoted error below $0.5$ rad and both intensities entering the phase difference above twice their errors.  Open: the quoted error is below $0.5$ rad, but at least one of the two intensities lies within twice its error of zero. Grey: the quoted error is $0.5$ rad or more. Crosses on the zero line mark the bins below the $K\bar K$ threshold in which BESIII constrained the three $2^{++}$ phases to be equal. The solid curve with its $68\%$ band is the fit, which uses only the two
$M2$ phase sets.}
\label{fig:phases}
\end{figure*}

Once the real production constants are fixed, the relative phases between the multipoles follow from the Omn\`es matrix and the kinematic constraints. Fits to the intensities admit competing solutions with different relative signs of the $h_2$ and $h_3$ contributions. These sign families have intensity $\chi^2$ values spanning a range of about $40$. The extracted resonance widths for the families differs by about $10\ \mathrm{MeV}$.

In $\pi^0\pi^0$ the signs follow by continuity from the Watson region: below the $K\bar K$ threshold all three $\pi\pi$ multipoles carry the same Omn\`es phase, so that $h_{M2}/h_{E1}$ and $h_{E3}/h_{E1}$ are real, and we choose both ratios to be positive. This is also the sign choice adopted by the BESIII fit~\cite{BESIII:2015rug}. The same sign choice follows from the  $h_2=h_3=0$ limit of Sec.~\ref{sec:mult}, in which $h_{M2}/h_{E1}>0$ and $h_{E3}/h_{E1}>0$ for all $s<q^2$.  Since the data follow this limit closely (see Fig.~\ref{fig:hierarchy}), a small $h_2$ or $h_3$ admixture cannot reverse the sign. For $M2$ the data confirm this sign choice: the measured $\pi^0\pi^0$ $M2$-$E1$ phase sits at zero throughout the $f_2(1270)$ resonance.  No such argument is available for $K_SK_S$, and there the phase data determine the sign.  At the $f_2'(1525)$ peak both amplitudes are dominated by the same pole and their relative phase is the sign of a ratio of couplings. The measured phase sits near zero, while the families with $M2$ in anti-phase in $K_SK_S$ are excluded, with $\chi^2/N$ values of about $300$.  One of these families gives the lowest $\chi^2$ to the intensities and the hadronic data alone, and would therefore be adopted without the phase data. Table~\ref{tab:poles} lists its pole parameters next to those of the adopted fit: including the phase data shifts both widths by more than their statistical errors while leaving the masses unchanged.

The sign of $E3$ in $\pi^0\pi^0$ is based on the Watson-region argument and the phase data.  Below the $K\bar K$ threshold, Watson's theorem fixes the $E3$-$E1$ phase to be either zero or $\pm\pi$. The positive sign is consistent with the $14$ credible $E3$-$E1$  phases above the $K\bar{K}$ threshold, all within $1.1$ rad of zero and none near $\pi$. These phases are not included in the fit and serve as a check on the model. The agreement is only partial: they are described with $\chi^2/N=13.6$ and lie above the predicted curve by $0.3$ rad on average. This is the only place where the model visibly falls short of the data, and it concerns an amplitude that carries only $3\%$ of the intensity. The opposite $E3$ sign gives $\chi^2/N$ above $100$ for these phases and is rejected.  The corresponding sign family is nevertheless refitted as one of the variations considered in Appendix~\ref{app:syst}, so that its effect on the pole parameters is included in the systematic uncertainty. The credible $E3$ phases thus help select the sign family, but they remain poorly described. The two model tests of Secs.~\ref{sec:kmat} and~\ref{sec:mult} do not improve their description. The phases of the Omn\`es matrix and of the individual multipoles, from which these differences arise, are shown in Appendix~\ref{app:cont}.

\subsection{The BESIII solution ambiguity}
\label{sec:branch}

A mass-independent partial-wave analysis of BESIII~\cite{BESIII:2015rug,BESIII:2018ubj} does not determine the amplitude uniquely. In each mass bin the amplitude is a polynomial in $\cos\theta$, and replacing one of its complex zeros by the complex conjugate changes the partial waves but leaves the angular distribution unchanged~\cite{Barrelet:1971pw}.  BESIII~\cite{BESIII:2015rug,BESIII:2018ubj} publish the resulting partner solution alongside the nominal one. We use the nominal solution~1 in both final states, and the fit supports this choice. Repeating the complete analysis with the partner solution increases $\chi^2$, with $\Delta\chi^2=+764$ for $\pi^0\pi^0$ partner and $+250$ for $K_SK_S$ partner. The fit cannot reproduce the partner solutions' $M2$ phases with real subtraction constants, and this accounts for most of the increase in $\chi^2$.

\subsection{Pole parameters}
\label{sec:poles}

\begin{table*}[!tbp]
\caption{Pole parameters of the two tensor mesons, $\sqrt{s_p}=M_p-i\Gamma_p/2$ in MeV, on the Riemann sheet~III ($\mathrm{Im}\,p_1<0$ and $\mathrm{Im}\,p_2<0$).  The first three rows are this work, with statistical uncertainties (bootstrap, with the errors scaled to the observed scatter) and, for the first row, systematic ones (see Table~\ref{tab:syst}).  The PDG entry for the $f_2(1270)$ is its $T$-matrix pole estimate.  For the $f_2'(1525)$ the PDG gives a Breit-Wigner mass and width, which need not coincide with the pole.}
\label{tab:poles}
\begin{ruledtabular}
\begin{tabular}{lllll}
 & \multicolumn{2}{l}{$f_2(1270)$} & \multicolumn{2}{l}{$f_2'(1525)$} \\
 & $M_p$ & $\Gamma_p$ & $M_p$ & $\Gamma_p$ \\ \hline
this work, all multipoles and $M2$ phases & $1267.5\pm1.3\pm2.6$ & $175.5\pm2.4\pm7.4$ & $1514.8\pm2.8\pm1.5$ & $80.4\pm3.4\pm4.8$ \\
this work, all multipoles, no phases & $1267.8\pm1.4$ & $180.5\pm2.1$ & $1512.5\pm2.5$ & $74.6\pm3.1$ \\
this work, $E1$ spectra only & $1269.9\pm1.3$ & $183.1\pm2.6$ & $1510.9\pm3.5$ & $72.1\pm4.5$ \\
\hline
Ref.~\cite{Rodas:2021tyb}, $N/D$, $J/\psi\to\gamma\pi\pi,\,\gamma K\bar K$ & $1268\pm8$ & $201\pm11$ & $1503\pm11$ & $84\pm15$ \\
Ref.~\cite{BESIII:2025dfb}, $K$-matrix, $\psi(3686)\to\gamma K_SK_S$ & $1226.2\pm8.6\pm8.8$ & $157.8\pm15.0\pm23.4$ & $1515.7\pm4.3\pm3.3$ & $67.4\pm7.6\pm8.2$ \\
Ref.~\cite{CrystalBarrel:2019zqh}, $K$-matrix, $\bar pp$ and scattering data & $1263.3\pm0.2\pm1.5$ & $193.7\pm0.4\pm1.6$ & $1495.0\pm1.1\pm8.1$ & $104.8\pm0.9\pm9.8$ \\
Ref.~\cite{Pelaez:2025jrn}, dispersive, $\pi\pi$ scattering & $1265.7^{+0.4}_{-0.5}$ & $197.2^{+1.0}_{-0.8}$ & - & - \\
PDG~\cite{ParticleDataGroup:2026mpi}, pole estimate / Breit-Wigner & $1260$-$1283$ & $180$-$220$ & $1517.3\pm2.4$ & $72^{+7}_{-6}$ \\
\end{tabular}
\end{ruledtabular}
\end{table*}

The poles are found on the Riemann sheet~III, which corresponds to
$\mathrm{Im}\,p_1<0$ and $\mathrm{Im}\,p_2<0$.  Writing $\sqrt{s_p}=M_p-i\Gamma_p/2$, the fit gives
\begin{align}
  \sqrt{s_p}(f_2(1270))&=(1267.5\pm1.3\pm2.6)\nonumber\\
    &\quad-i\,\frac{1}{2}\,(175.5\pm2.4\pm7.4)\ \mathrm{MeV},\nonumber\\
  \sqrt{s_p}(f_2'(1525))&=(1514.8\pm2.8\pm1.5)\nonumber\\
    &\quad-i\,\frac{1}{2}\,(80.4\pm3.4\pm4.8)\ \mathrm{MeV},
  \label{eq:poles}
\end{align}
The first uncertainty is statistical and the second systematic, as described in Sec.~\ref{sec:err}. The residues show the near-ideal mixing pattern, the $f_2(1270)$ coupling dominantly to $\pi\pi$ and the $f_2'(1525)$ to $K\bar K$. This pattern is not imposed in the fit. We do not quote residue branching fractions, because with two channels they are normalised to $\pi\pi$ and $K\bar K$ alone.

Additionally, Table~\ref{tab:poles} lists the results of the fit without the phase differences data and the fit restricted to the two $E1$ spectra. All three fits use the same hadronic model and twelve-constant production sector. They differ only in the data included.  Compared with the $E1$-only fit, the inclusion of the subleading multipoles and the phase data reduces the statistical uncertainties by factors of $1.1$ to $1.5$, with the errors quoted by the experiments in both cases, the largest gain being on the $f_2(1270)$ width.  Including these data also lowers the $f_2(1270)$ width by $7.7\ \mathrm{MeV}$, corresponding to $3.0$ standard deviations in the $E1$-only fit.  The $M2$ and $E3$ spectra are thus not redundant with the $E1$ spectrum and bring extra information.

The remaining rows of Table~\ref{tab:poles} are other pole determinations we are aware of. In Ref.~\cite{Rodas:2021tyb}, the authors fit the BESIII $0^{++}$ and $2^{++}$ $E1$ waves of both final states within the $N/D$ formalism, without the scattering data.  The BESIII analysis of $\psi(3686)\to\gamma K_SK_S$~\cite{BESIII:2025dfb} uses a one-channel $K$-matrix with three tensor poles and provides the $T$-matrix pole listed by the PDG for the $f_2'(1525)$.  The Crystal Barrel analysis~\cite{CrystalBarrel:2019zqh} is a four-channel $K$-matrix fit to $\bar pp$ annihilation data at $900\ \mathrm{MeV}/c$ together with the $\pi\pi$, $K\bar K$ and $\eta\eta$ scattering data.  For the $f_2(1270)$ alone, Ref.~\cite{Pelaez:2025jrn} gives a parametrisation-independent pole from the analytic continuation of forward dispersion relations, using the global dispersive analysis of Ref.~\cite{Pelaez:2024uav} as input.  Its three global fits give widths between $182$ and $197\ \mathrm{MeV}$ (in Table~\ref{tab:poles} we quote the authors' preferred solution).

The $f_2(1270)$ mass agrees with all of these determinations to within a few MeV, apart from the low value of the $\psi(3686)$ analysis. The width of the $f_2(1270)$ is where the determinations spread. Our result, $175.5\pm2.4\pm7.4\ \mathrm{MeV}$, is at the low end: $7\ \mathrm{MeV}$ below the two alternative dispersive solutions of Ref.~\cite{Pelaez:2025jrn}, $22\ \mathrm{MeV}$ below their preferred one. It also lies $18\ \mathrm{MeV}$ below Crystal Barrel and $26\ \mathrm{MeV}$ below JPAC, and $4.5\ \mathrm{MeV}$ below the lower edge of the PDG pole range.  Only the $\psi(3686)$ analysis' value, with its large errors, is located lower.  We return to this in Sec.~\ref{sec:err}. For the $f_2'(1525)$, the mass is determined to a few MeV and agrees with the PDG Breit-Wigner value within $0.7\sigma$ and with the $\psi(3686)$ analysis' pole within $0.1\sigma$, whereas the width is the least-known quantity.  Our pole width, $80.4\pm3.4\pm4.8\ \mathrm{MeV}$, lies about $1\sigma$ above both the PDG value and the $\psi(3686)$ pole width, in units of the combined uncertainties, while it agrees with JPAC.  The Crystal Barrel pole stands out from all the others, with a larger width and lower mass.

\subsection{Uncertainties}
\label{sec:err}

The statistical uncertainties come from one parametric bootstrap with $400$ replicas.  Each replica is generated by applying a Gaussian fluctuation to every fitted data point (hadronic data, intensities and phase differences) using its assigned error as the standard deviation. The complete model, hadronic parameters and production constants, is then refitted starting from the nominal solution.  The quoted uncertainty is half the difference between the $16$th and $84$th  percentiles. Several sets are not described within the errors quoted by the experiments (see Table~\ref{tab:chi2}), so before generating the replicas, the errors of each data set with $\chi^2/N>1$ are multiplied by $S=\sqrt{\chi^2/N}$, following the PDG prescription for inconsistent inputs~\cite{ParticleDataGroup:2026mpi}. The factors are listed in the last column of Table~\ref{tab:chi2}.  The scaling increases the uncertainty in the $f_2(1270)$ width from $\pm1.7$ to $\pm2.4\ \mathrm{MeV}$ and changes the other uncertainties by at most $0.2\ \mathrm{MeV}$. After the refit, no replica had a different sign family from the nominal solution. The two other ensembles of Table~\ref{tab:poles} are generated in the same way, the fit without phases with its own scale factors, and the $E1$-only fit uses the errors as quoted.

The systematic uncertainties are estimated from the $17$ variations in Appendix~\ref{app:syst}, each involving a complete refit with one ingredient of the analysis changed. They include the two model tests of Secs.~\ref{sec:kmat} and~\ref{sec:mult}, the general background and the quadratic production polynomials. Both use exactly the same inputs as the adopted fit: the data of Table~\ref{tab:chi2}, the phase cuts, the kinematic constraints and real production constants. Their pole shifts thus enter the systematic uncertainty, separately from the bootstrap errors. The largest absolute shift in each column is the systematic uncertainty quoted for the pole positions in Eq.~\eqref{eq:poles}: $2.6\ \mathrm{MeV}$ on the $f_2(1270)$ mass (Hyams phase shift removed), $7.4\ \mathrm{MeV}$ on its width (quadratic production polynomials), $1.5\ \mathrm{MeV}$ on the $f_2'(1525)$ mass (constant instead of linear polynomial for $h_3$) and $4.8\ \mathrm{MeV}$ on its width (constant instead of linear polynomial for $h_3$). The $f_2'(1525)$ width is fixed mainly by the $J/\psi\to\gamma K_SK_S$ spectra. It still moves by up to $4\ \mathrm{MeV}$ when the hadronic inputs are varied, because the two resonances share the $\pi\pi\to K\bar K$ amplitude of the $K$-matrix.

Neither the statistical nor the systematic uncertainty accounts for the two-channel truncation.  The $f_2(1270)$ decays to $4\pi$ with a branching fraction of about $10\%$, twice its $K\bar K$ fraction~\cite{ParticleDataGroup:2026mpi}, and the $\pi\pi$ $D$ wave is inelastic from about $0.9\ \mathrm{GeV}$ onward~\cite{Pelaez:2024uav,Pelaez:2025jrn}.  The two-channel amplitude is elastic below the $K\bar K$ threshold and compensates for the omitted channels through its fitted couplings.  The effect is visible in the $f_2(1270)$ width: the Hyams phase shift data above $0.9\ \mathrm{GeV}$ are the only hadronic data set not described within its errors. Our width lies at the lower of the determinations in Table~\ref{tab:poles}, $7$ to $22\ \mathrm{MeV}$ below the dispersive values of Ref.~\cite{Pelaez:2025jrn}, which include the inelasticity. Every variation of Table~\ref{tab:syst} retains the two-channel truncation, so neither their spread nor the bootstrap can quantify its effect.  We therefore regard the $f_2(1270)$ width of Eq.~\eqref{eq:poles} as carrying an additional model uncertainty of up to $20\ \mathrm{MeV}$ upward, which is not included in the quoted errors. Addressing it would require a third effective channel constrained by the measured inelasticity. This would change the asymptotic condition of the Omn\`es problem and is beyond the scope of the present work. The $f_2'(1525)$ is not visibly affected: its $\eta\eta$ mode, accounting for about $10\%$ of the width, opens $100\ \mathrm{MeV}$ above the $K\bar K$ threshold, where the two-channel amplitude is unitary, and the resonance is constrained directly by the $K_SK_S$ spectra. 

\section{Summary}

We have determined the pole parameters of $f_2(1270)$ and $f_2'(1525)$ from one coupled-channel fit to the isoscalar $D$-wave $\pi\pi$ phase shift, the $\pi\pi\to K\bar K$ modulus, the three $2^{++}$ radiative multipoles of $J/\psi\to\gamma\pi^0\pi^0$ and $J/\psi\to\gamma K_SK_S$, and the $M2$-$E1$ phases where BESIII determines them, without any resonance input from the PDG. The hadronic amplitude is a two-channel $K$-matrix with two poles.  The radiative decays enter through the coupled-channel Omn\`es matrix, rebuilt from it at every step, and the kinematic constraints among the $J=2$ helicity amplitudes relate the six spectra to three regular functions per channel.  The fit describes $358$ points with $\chi^2/N=1.29$ and gives the pole positions of Eq.~\eqref{eq:poles}.

Three points deserve emphasis.  First, the multipole intensities alone do not fix the relative signs of the three regular functions $h_1$, $h_2$ and $h_3$, which are the partial-wave amplitudes free of kinematic constraints. The resulting sign families differ in the resonance widths by about $10\ \mathrm{MeV}$. In the $K_SK_S$ channel, the sign is fixed by the $M2$-$E1$ phase measured by BESIII on the $f_2'(1525)$ peak. In the $\pi^0\pi^0$ channel, Watson's theorem makes the multipole ratios real below the $K\bar K$ threshold, so that only their sign remains open. This sign is positive in the $h_2=h_3=0$ limit, which the measured intensity ratios follow closely, and the measured $M2$-$E1$ phases confirm it independently. The credible $E3$ phases also help select the sign family, but they remain poorly described.

Second, the $f_2'(1525)$ mass agrees with the PDG value~\cite{ParticleDataGroup:2026mpi} and with the $\psi(3686)$ analysis' pole determination to a few MeV. Its width, which is less well known than its mass, is obtained as $80.4\pm3.4\pm4.8\ \mathrm{MeV}$. This is about $1\sigma$ above the PDG fit value and compatible with the JPAC pole~\cite{Rodas:2021tyb}.

Third, the $f_2(1270)$ width, $175.5\pm2.4\pm7.4\ \mathrm{MeV}$, falls below the dispersive determinations. We attribute this difference to the neglected $4\pi$ inelasticity rather than to the data. This effect is not included in the quoted errors and is estimated to shift the width upward by up to $20\ \mathrm{MeV}$.

The same construction applies to any radiative decay in which the $J=2$ multipoles have been separated. The Omn\`es matrix obtained here is available as dispersive input for the isoscalar $D$-wave $\pi\pi/K\bar K$ final-state interaction.

\section*{Acknowledgements}

This work was supported by the Deutsche Forschungsgemeinschaft (DFG, German Research Foundation) within the Research Unit [Photon-photon interactions in the Standard Model and beyond, Projektnummer 458854507 - FOR 5327]. The work of EP is supported by the Spanish Government (Agencia Estatal de Investigaci\'on MCIN/AEI/10.13039/501100011033) Grants No. PID2023-146220NB-I00, and CEX2023-001292-S (Agencia Estatal de Investigaci\'on MCIU/AEI (Spain) under grant IFIC Centro de Excelencia Severo Ochoa), by the Generalitat Valenciana (Spain) through the plan GenT program CIDEGENT/2021/037, by the U.S. National Science Foundation under grant PHY-2310149 and by the U.S. Department of Energy ExoHad Topical Collaboration, contract DE-SC0023598. We acknowledge the use of Claude Fable 5.1, 5.2 (Anthropic) and GPT Astra (OpenAI) for cross-checks of our results and for assistance with the code. All results were verified independently by the authors, who take full responsibility for the content of this work.

\appendix

\section{Phases in the continuous convention}
\label{app:cont}

\begin{figure*}[!tbp]
\includegraphics[width=0.84\textwidth]{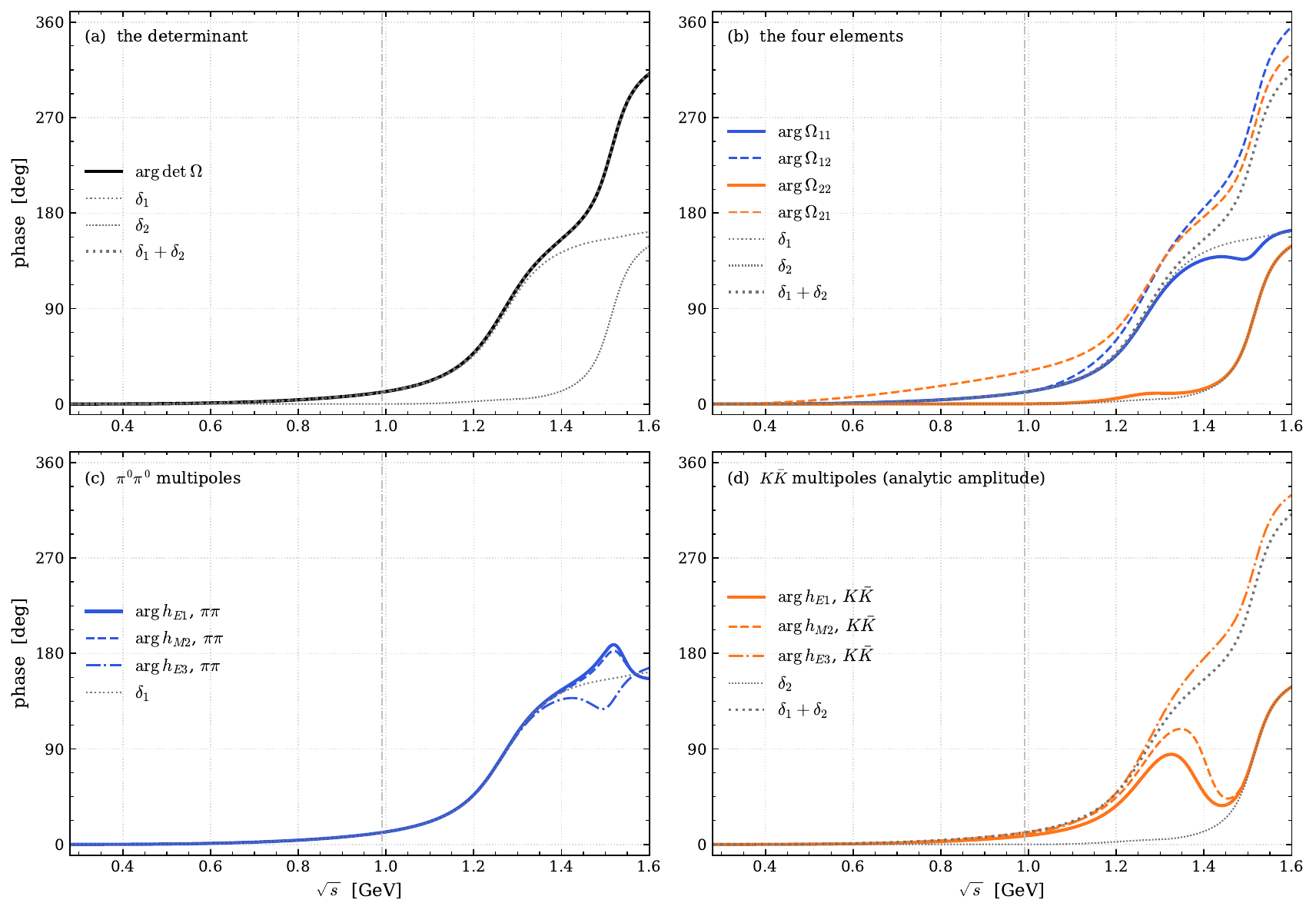}
\caption{Phases in the continuous convention: unwrapped from the $\pi\pi$ threshold, where every amplitude is real, with the sign of the real threshold value divided out, so that all of them start at zero. (a) The phase of $\det\bm{\Omega}$, which equals $\delta_1+\delta_2$ identically.  (b) The four elements of the Omn\`es matrix.  (c) The three $\pi^0\pi^0$ multipoles against $\delta_1$.  (d) The three $K\bar K$ multipoles with the threshold factor $(s-4m_K^2)$ divided out, so that their phases are continuous through the threshold, against $\delta_2$ and $\delta_1+\delta_2$.  The dash-dotted line marks the $K\bar K$ threshold.}
\label{fig:continuous}
\end{figure*}

Figure~\ref{fig:continuous} shows where the phase differences discussed in Sec.~\ref{sec:phases} come from, with every phase unwrapped from the $\pi\pi$ threshold such that it starts at zero and stays continuous. In $\pi^0\pi^0$ all three multipoles carry $\delta_1$ below the $K\bar K$ threshold and stay within about $30^\circ$ of it above the threshold, explaining the small tabulated $\pi^0\pi^0$ phase differences. In $K\bar K$ the $E1$ and $M2$ amplitudes follow $\delta_2$ at and above the $f_2'(1525)$, while $E3$, fed by the $\pi\pi\to K\bar K$ element $\Omega_{21}$, follows $\delta_1+\delta_2$. The $E3$-$E1$ relative phase is therefore close to $\delta_1$. The predicted rise of $\arg h_{E3}-\arg h_{E1}$ through $\pi$ across $1.2$-$1.5\ \mathrm{GeV}$ in $K_SK_S$ reflects the $f_2(1270)$ contribution to the kaon channel from the coupled-channel dynamics, not the $f_2'(1525)$. BESIII cannot determine this phase, but the direction of its variation differs between the two $E3$ signs. A measurement with higher statistics would settle the question.

\section{Systematic variations}
\label{app:syst}

\begin{table*}[!tbp]
\caption{Systematic variations of the analysis and the resulting shifts of the pole parameters, in MeV, with respect to the adopted fit.  Every row is a complete refit of the hadronic parameters and the production constants.  The largest absolute shift in each column is the systematic uncertainty quoted for the pole positions in Eq.~\eqref{eq:poles}.}
\label{tab:syst}
\begin{ruledtabular}
\begin{tabular}{lrrrr}
variation & $\Delta M_p(f_2(1270))$ & $\Delta\Gamma_p(f_2(1270))$ & $\Delta M_p(f_2'(1525))$ & $\Delta\Gamma_p(f_2'(1525))$ \\ \hline
errors rescaled by $\sqrt{\chi^2/N}$ & $-0.3$ & $-0.0$ & $-0.3$ & $-0.8$ \\
CFD errors doubled & $-0.3$ & $-0.6$ & $-0.0$ & $+1.0$ \\
CFD points removed & $-1.6$ & $-1.5$ & $-0.4$ & $+4.0$ \\
Hyams phase shift only up to $1.3\ \mathrm{GeV}$ & $-0.8$ & $-2.3$ & $-0.1$ & $-1.0$ \\
Hyams phase shift only below the $K\bar K$ threshold & $-2.5$ & $-2.7$ & $-0.3$ & $-1.1$ \\
Hyams phase shift removed & $-2.6$ & $-2.6$ & $-0.3$ & $-1.1$ \\
$\pi\pi\to K\bar K$ errors doubled & $-0.5$ & $-0.5$ & $+0.0$ & $+3.2$ \\
BESIII window $1.0$-$1.6\ \mathrm{GeV}$ & $+1.6$ & $+2.2$ & $+0.2$ & $-0.4$ \\
BESIII window $0.9$-$1.55\ \mathrm{GeV}$ & $-0.0$ & $+0.1$ & $-1.3$ & $-3.7$ \\
BESIII window $0.9$-$1.65\ \mathrm{GeV}$ & $-0.2$ & $+0.0$ & $-1.5$ & $+1.1$ \\
phase credibility: quoted error below $0.35$ rad & $+0.0$ & $+0.0$ & $+0.0$ & $+0.0$ \\
phase credibility: amplitudes above $3\sigma$ & $+0.0$ & $+0.0$ & $+0.0$ & $+0.0$ \\
$\pi^0\pi^0$ $E3$ phases fitted, errors doubled & $+0.0$ & $+0.0$ & $+0.0$ & $+0.0$ \\
constant instead of linear polynomial for $h_3$ & $-0.2$ & $-0.7$ & $-1.5$ & $+4.8$ \\
quadratic production polynomials & $+0.8$ & $+7.4$ & $-0.3$ & $-3.4$ \\
background $C_{ab}b_ab_b/(s_B-s)$, $s_B=-0.5\ \mathrm{GeV}^2$ & $-1.6$ & $-0.6$ & $+1.2$ & $-1.2$ \\
sign family with $E3$ in anti-phase & $+0.4$ & $+2.5$ & $-0.5$ & $+1.1$ \\
\end{tabular}
\end{ruledtabular}
\end{table*}

Table~\ref{tab:syst} lists the variations used to estimate the systematic uncertainties in Eq.~\eqref{eq:poles}.  Each row corresponds to a complete refit with one ingredient changed. The variations cover the error rescaling, the weight or range of each hadronic input, the BESIII fit window, the credibility cuts on the phases, the inclusion of the $E3$ phases, a constant instead of a linear polynomial for $h_3$, quadratic production polynomials and the general background of Sec.~\ref{sec:kmat}. Lastly, they also include the sign family with $E3$ in anti-phase, which is the only other family compatible with the $M2$ phases.

\bibliography{references}

\end{document}